\documentclass[usenatbib]{mn2e}
\usepackage{graphics,longtable}
\usepackage{txfonts}
\title[PG1115+080: A2/A1 flux ratio and time delays]
{PG 1115+080: variations of the A2/A1 flux ratio and new values of the time delays}
\author [Tsvetkova V.S. et al.] { \parbox[t]{\textwidth}{
\vspace{-1.0cm} 
 V.S.Tsvetkova$^1$\thanks {E-mail: tsvetkova@astron.kharkov.ua}, 
 V.G.Vakulik$^{1,2}$,  V.M.Shulga$^1$, R.E.Schild$^3$,  V.N.Dudinov$^{1,2}$,\\
       A.A.Minakov$^1$, S.N.Nuritdinov$^5$, B.P.Artamonov$^5$, 
       A.Ye.Kochetov$^{1,2}$, G.V.Smirnov$^2$, A.A.Sergeyev$^{1,2}$, 
       V.V.Konichek$^2$, I.Ye.Sinelnikov$^2$, A.P.Zheleznyak$^2$, 
       V.V.Bruevich$^4$, R.Gaysin$^5$, T.Akhunov$^5$, O.Burkhonov$^5$}
       \vspace*{6pt}\\
       $^1$Institute of Radio Astronomy of Nat.Ac.Sci. of Ukraine,
           Krasnoznamennaya 4, 61002 Kharkov, Ukraine\\
       $^2$Institute of Astronomy of Kharkov National University, Sumskaya
           35, 61022 Kharkov, Ukraine\\
       $^3$Center for Astrophysics, 60 Garden Street, Cambridge, MA
           02138, U.S.A.\\
       $^4$Sternberg Astronomical Institute, Universitetski Ave. 13, 119899 Moscow,
            Russia\\  
       $^5$Astronomical Institute of Ac.Sci.of Uzbekistan, Astronomicheskaya 33,
           100052 Tashkent, Uzbekistan\\
        \vspace*{-0.5cm}}          
          
\begin{document}     
\date{Accepted ...
      Received ...;
      in original form December 24, 2009}

\pagerange{\pageref{firstpage}--\pageref{lastpage}}
\pubyear{...}

\maketitle

\label{firstpage}

\begin{abstract}
We report the results of our multicolor observations of PG 1115+080 with the 
1.5-m telescope of the Maidanak Observatory (Uzbekistan, Central Asia) in 
2001-2006. Monitoring data in filter $R$ spanning the 2004, 2005 and 2006 
seasons (76 data points) demonstrate distinct brightness variations of the 
source quasar with the total amplitude of almost 0.4 mag. Our $R$ light curves 
have shown image C leading B by 16.4d and image (A1+A2) by 12d that is 
inconsistent with the previous estimates obtained by Schechter et al. in 1997 
-- 24.7d between B and C and 9.4d between (A1+A2) and C. The new values of 
time delays in PG 1115+080 must result in larger values for the Hubble 
constant, thus reducing difference between its estimates taken from the 
gravitational lenses and with other methods. Also, we analyzed variability 
of the A2/A1 flux ratio, as well as color changes in the archetypal "fold" 
lens PG 1115+080. We found the A1/A2 flux ratio to grow during 2001-2006 
and to be larger at longer wavelengths. In particular, the A2/A1 flux ratio 
reached 0.85 in filter $I$ in 2006. We also present evidence that both the 
A1 and A2 images might have undergone microlensing during 2001-2006, with 
the descending phase for A1 and initial phase for A2. We find that the A2/A1 
flux ratio anomaly in PG 1115 can be well explained both by microlensing and 
by finite distance of the source quasar from the caustic fold.

\end{abstract}

\begin{keywords}
cosmology: gravitational lensing -- galaxies: quasars: individual:
PG 1115+080.
\end{keywords}

\section{Introduction}

Gravitationally lensed quasars are known to potentially provide estimates 
of the Hubble constant $H_0$ from measurements of the time delays between the 
quasar intrinsic brightness variations seen in different quasar images (Refsdal 
1964).  Since a phenomenon of gravitational lensing is controlled
by the surface density of the total matter (dark plus luminous), it provides a 
unique possibility both to determine the value of $H_0$ and to probe the dark 
matter content in lensing galaxies and along the light paths in the medium 
between the quasar and observer. 

By now the time delays have been measured in about 20 gravitationally lensed
quasars resulting in the values of $H_0$ that are generally noticeably
less than the most recent estimate of $H_0$ obtained in the HST Hubble
Constant Key Project with the use of Cepheids -- $H_0=72\pm8$ km s$^{-1}$
Mpc$^{-1}$ (Freedman et al. 2001). This discrepancy is large enough and, if
the Hubble constant is really a universal constant, needs to be explained.
A detailed analysis of the problem of divergent $H_0$ estimates inherent in
the time delay method, and the ways to solve it can be found, e.g., in
Keeton \& Kochanek (1997), Saha \& Williams (1997), Kochanek (2002), Kochanek 
\& Schechter (2004) and Schechter (2005), Read et al. (2007), and in many 
other works.

The main sources of uncertainties in determining $H_0$ are:
\begin{itemize}
\item low accuracy of the time delay estimates caused by poorly sampled and
insufficiently accurate light curves of quasar components, as well as by
microlensing events and, as a rule, by low amplitudes of the quasar intrinsic
variability;
\item difference in the values of cosmological constants adopted in deriving
$H_0$;
\item invalid models of mass distribution in lensing galaxies.
\end{itemize}

The way to reduce the effect of the first source of errors is clear enough:
more accurate and better sampled light curves of a sufficient duration are
needed. A choice of the cosmological model is usually just indicated - this
is mostly a question of agreement. As to the third item, here the problem of
estimating the Hubble constant encounters the problem of the dark matter
abundance in lensing galaxies. 

The problem of determining the Hubble constant from the time delay lenses is
known to suffer from the so-called central concentration degeneracy, which
means that, given the measured time delay values, the estimates of the Hubble
constant turn out to be strongly model-dependent. In particular, models with
more centrally concentrated mass distribution (lower dark matter content)
provide higher values of $H_0$, more consistent with the results of the local
$H_0$ measurements than those with lower mass concentration towards the
center (more dark matter). Moreover, it has long been noticed that the time 
delays are sensitive not only to the total radial mass profiles of lensing 
galaxies, but also to the small perturbations in the lensing potential (e.g., 
Blandford \& Narayan 1986,  Witt et al. 2000, Oguri 2007). It is interesting 
to note that this effect has been recently proposed as a new approach to detect 
dark matter substructures in lensing galaxies (Keeton \& Moustakas 2009). 

The Hubble constant -- central concentration degeneracy is a part of the well 
known total problem of lensing degeneracies: all the lensing observables, even if
they were determined with zero errors, are consistent with a variety of the mass
distribution laws in lensing galaxies. A strategy for solving this non-uniqueness
problem could be a search through a family of lens models that are capable of 
reproducing the lensing observables (Williams \& Saha 2000, Oguri et al. 2007). 
Then many models can be run in order to infer a probability density for a 
parameter under investigation, e.g. for $H_0$ (Williams \& Saha 2000). The most 
recent studies (Saha et al. 2006, Read et al. 2007) have shown that, in such an 
approach, the discrepancy between the $H_0$ value determined from lensing and 
with other methods can be substantially reduced if non parametric models for 
mass reconstruction are used, which can provide much broader range of models as 
compared to the parametric ones.

In defining priors on the allowed space of lens models, it is naturally to 
assume that lensing galaxies in the time delay lenses are similar in their mass 
profiles to other early-type ellipticals, that are presently believed to be 
close to isothermal and admit the presence of the cold dark matter haloes. 
The isothermal models are also consistent with stellar dynamics, as well as 
with the effects of strong and weak lensing. 

The quadruply lensed quasars are known to be more promising for solving
these problems as compared to the two-image lenses since they provide
more observational constraints to fit the lens model. Ten astrometric
constraints can be presently regarded as measured accurately enough for
most systems. This especially concerns the relative coordinates of quasar
images. As to the lensing galaxiy, its less accurate coordinates are 
often the only reliable information about the lensing object known from 
observations, with other important characteristics being derived indirectly. 
This situation is inherent, e.g. in PG 1115+080 with its faint, 0.31-
redshift galaxy. Of other observational constraints, the time delays 
and their ratios are very important. In quadruple lenses, the time delay 
between one of the image pairs is usually used to determine $H_0$, while 
the other ones form the $H_0$-independent time delay ratios to constrain 
the lens model, (Keeton \& Kochanek 1997). 

It has long been known that the observed positions of multiple quasar
macroimages are well predicted by smooth regular models of mass distribution
in lensing galaxies, while their brightness ratios are reproduced by such
models poorly, (e.g., Kent and Falco 1988, Kochanek 1991, Mao \& Schneider 
1998). The first systematic analysis of this problem called "flux ratio 
anomalies" was made by Mao \& Schneider (1998), who assumed that the 
anomalies of mutual fluxes of the components in some lenses can be explained 
by the presence of small-scale structures (substructures) in lensing galaxies or somewhere near the line of sight.

A popular model of forming hierarchical structures in the Universe with a
dominant content of dark matter is currently known to poorly explain the
observed distribution of matter at small scales. In particular, the
expected number of satellite galaxies with masses of the order of
$M_G\approx10^8M_\odot$ remained after the process of hierarchical
formation is completed, is an order of magnitude larger than a number
of dwarf galaxies with such masses actually observed within the Local
Group (Klypin et al. 1999, Moore et al. 2001). One of the solutions
of this contradiction is a suggestion that some substructures, especially
those with low masses, are not luminous.
\begin{figure}
\centering
\resizebox{0.7\hsize}{!}{\includegraphics{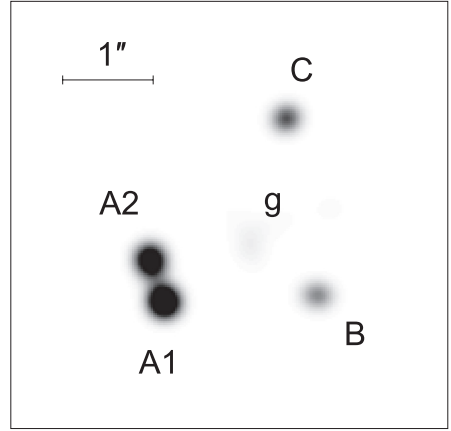}}
 \caption{PG1115+080 from observations in filter $R$ with the 1.5-m
 telescope of the Maidanak Observatory. The image was obtained by averaging of
 six frames from a series obtained in February 24, 2004, with a subsequent
 Richardson-Lucy processing.  }
\end{figure}

Metcalf and Madau (2001) were the first to note that the dark matter
paradigm can naturally explain existence of substructures in galaxies
lensing the remote quasars, as proposed by
Mao and Schneider (1998) to interpret the anomalies of mutual fluxes
of quasar macroimages, and vice versa, confirmation of substructures
with masses from $10^6M_\odot$ to $10^8M_\odot$ is capable of removing
the contradiction between the predicted number of the low-mass satellite
galaxies and that one actually observed. The idea turned out to be
intriguing and was immediately taken up, (Brada\v c et al. 2002, Chiba
2002, Dalal \& Kochanek 2002, Keeton 2001, Metcalf \& Zhao 2002).
Investigation of flux ratio anomalies in gravitationally lensed quasars is
presently believed to be a powerful tool in solving the problem of
 the dark matter abundance in the Universe. It is intensively discussed in
numerous recent publications (Congdon \& Keeton 2005; Keeton et
al. 2003, 2005; Kochanek \& Dalal 2004; Mao et al. 2004; 
 Miranda \& Jetzer 2007; Pooley et al. 2006, 2007, 2009; Morgan et 
al. 2008).

\begin{figure*}
\centering
\resizebox{0.8\hsize}{!}{\includegraphics{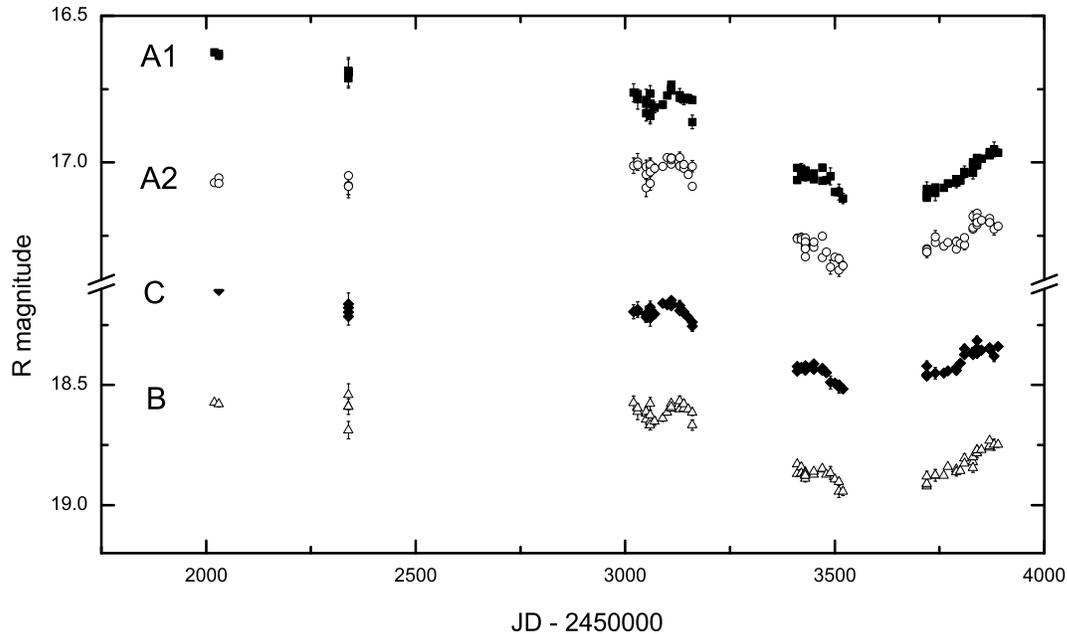}}
 \caption{The light curves of PG 1115+080 A1, A2, B, C from observations in
 filter $R$ with the 1.5-m telescope of the Maidanak Observatory in 2001,
 2002, 2004, 2005 and 2006.}
\end{figure*}

In Sec. 3 we report our measurements of the A2/A1 flux ratios in filters 
$V$, $R$ and $I$ from our data obtained in 2001, 2002 and 2004-2006 at the 
Maidanak Observatory and analyse their behaviors in time and in wavelength. 
In Sec. 4, we analyze the new estimates of the time delays between the PG 
1115+080 images, obtained from our monitoring in the $R$ filter during 
2004-2006 and reported in Vakulik et al. (2009). The new values differ 
noticeably from those reported by Schecter et al. (1997) and Barkana (1997). 
In Sec. 5, we discuss our results and their possible effect on selecting an 
adequate lens model for PG 1115+080 and estimating the value of the Hubble 
constant.

\section{Observations}
The quadruply imaged quasar PG 1115+080 is one of the most promising
candidates both to investigate the dark matter problem and to determine
the $H_0$ value from measurements of the time delays between the image
components. The source with a redshift  of $z_S=1.722$ is lensed by a galaxy
with $z_G=0.31$ (Henry \& Heasley 1986, Christian et al. 1987, Tonry 1998),
which forms four quasar images, with an image pair A1 and A2 bracketing the
critical curve very close to each other. It is the second gravitationally
lensed quasar discovered over a quarter of century ago, at first as a triple
quasar (Weymann et al. 1980). Hege et al. (1980) were the first to resolve
the brightest image component into two images separated by 0.48 arcsec.
Further observations (Young et al. 1981, Vanderriest et al. 1986, Christian
et al. 1987, Kristian et al. 1993, Courbin et al. 1997) have provided 
positions of quasar images and information about the lensing object, which  
allowed construction of a macrolens model (e.g., Keeton, Kochanek \& Seljak 
1997). In particular, Keeton \& Kochanek (1997) have shown that the observed 
quasar image positions and fluxes and the galaxy position can be fit well by 
an ellipsoidal galaxy with an external shear rather than by only an 
ellipsoidal galaxy, or by a circular galaxy with an external shear. They 
noted that a group of nearby galaxies detected by Young et al. (1981) could 
provide the needed external shear.

Observations of PG 1115+080 were started at the 1.5-meter telescope of the 
high-altitude Maidanak Observatory (Central Asia, Uzbekistan) in 2001. 
An image scale of 0.26\arcsec/pix was available at the f/8 focal plane with 
a scientific BroCam CCD camera having a SITe ST 005A 2030 x 800 chip. The 
CCD images were usually taken in series consisting of 2 to 10 frames for 
the $R$ filter and of 2 to 6 frames for $V$ and $I$. To provide higher 
photometric accuracy, we averaged the values of magnitudes estimated from 
individual frames. The seeing varied from 0.75\arcsec \,to 1.3\arcsec \,[the 
full width at half-maximum (FWHM) of images of the reference stars B and 
C according to designation by Vanderriest et al. (1986)]. The analysis of 
photometry shows no significant dependence of the photometry errors on 
seeing, excepting the FWHM noticably exceeding 1.3\arcsec. Occasional 
frames with such values of the FWHM were excluded from processing.

In Fig. 1 we show one of the best images of PG 1115+080 obtained through
the $R$ filter. For better view, the image was restored with an algorithm
similar to that proposed by Richardson (1972) in optics and independently 
by Lucy (1974) in astronomy (the Richardson-Lucy iterative method).

Our algorithm for photometric image processing is similar to that applied to
Q2237+0305 and described in detail by Vakulik et al. (2004). The 
light curves of PG 1115+080A1,A2,B,C in filter $R$ for the time period 
from April 2001 to June 2006 are shown in Fig. 2. The photometry in the 
$V$, $R$ and $I$ filters is presented in Tables 6-8.

Unfortunately, observations were not carried out in 2003, and the data are
very scanty for the 2001 and 2002 seasons in all the three filters. The most 
numerous data were obtained in filter $R$, especially in 2004 (23 nights),
2005 (27 nights) and in 2006 (24 nights). The data demonstrate noticeable 
variations of the quasar brightness, with the total amplitude reaching 
approximately 0.4 mag in 2004-2006, and smaller amplitudes of about 0.05 mag 
on a time-scale of two months, which are clearly seen in all the four light 
curves in 2004. The time delays between the light curves of the C and B, C 
and A1 (or A2) images can be easily seen from a simple visual inspection of 
the $R$ light curves, therefore, the data obtained in 2004-2006 in filter $R$ 
seem to have good prospects for obtaining reliable estimates of the time 
delays in PG 1115+080.

\begin{table*}
 \centering
  \caption{Estimates of the A2/A1 brightness ratios in PG 1115+030 for the 
  time period 1980-2006 from all available data. }
  \begin{tabular}{@{}lllll@{}}
  \hline
  Date  & A2/A1     & Spectral & Instrument &Reference \\
        & flux ratio& range    &            &  \\
  \hline
1980 June       & 0.83  & $V$ & MMT     & Hege et al. (1980)\\
1981 April 30   & 1.0   & $B$ & CFHT    & Vanderriest (1986)\\
1983 March 8    & 1.0   & $V$ & -- " -- & -- " -- \\
1984 March 26   & 0.95  & $B$ & -- " -- & -- " -- \\
1985 March 16   & 0.75  & $B$ & -- " -- & -- " -- \\ 
1985 March 19   & 0.79  & $V$ & -- " -- & -- " -- \\
1986 February 19& 0.79  & $V$ & CFHT    & Christian (1987) \\
-- " --         & 0.8   & $R$ & -- " -- & -- " -- \\
-- " --         & 0.79  & $B$ & -- " -- & -- " -- \\
1989 April      & 0.68  & $I$ & CFHT    & Schechter (1993) \\
1991 March 3    & 0.66  & $V$ & HST     & Kristian (1993) \\
-- " --         & 0.7   & $I$ & -- " -- & -- " -- \\
1992 April      & 0.67  & $I$ & Hiltner & Schechter 1993 \\
-- " --         & 0.69  & $V$ & -- " -- & -- " -- \\
-- " --         & 0.72  & $I$ & CTIO    & -- " -- \\
-- " --         & 0.68  & $V$ & -- " -- & -- " -- \\
1993 April      & 0.69  & $I$ & Hiltner & -- " -- \\
-- " --         & 0.63  & $V$ & -- " -- & -- " -- \\
1995 December 20& 0.66  & $V$ & Magellan& Pooley et al. (2006) \\
1996 June 7     & 0.68  & $I$ & NOT     & Courbin et al. (1997)\\
1997 November 17& 0.64  & $H$ & HST     & Impey et al. (1998)\\
?               & 0.52  & $V$ & HST     & Morgan et al. (2008)\\
?               & 0.67  & $I$ & -- " -- & -- " -- \\
?               & 0.63  & $H$ & -- " -- & -- " -- \\
2001 March 26   & 0.66  & $V$ & Magellan& Pooley et al. (2006)\\
2001 April 20-27& 0.74  & $V$ & 1.5m(Maidanak)& This work\\
-- " --         & 0.67  & $R$ & -- " -- & -- " -- \\
-- " --         & 0.72  & $I$ & -- " -- & -- " --\\
2002 March      & 0.76  & $V$ & -- " -- & -- " -- \\
-- " --         & 0.71  & $R$ & -- " -- & -- " -- \\
-- " --         & 0.72  & $I$ & -- " -- & -- " -- \\  
2004 February 22& 0.81  &Sloan $i^\prime$& Magellan& Pooley et al.(2006)\\
2004 May 5-6    & 0.93  &11.67$\mu$m& Subaru& Chiba et al. (2005)\\
2004 Jan.17-June 8&0.79 & $V$ & 1.5m Maidanak& This work\\
-- " --         & 0.81  & $R$ & -- " -- &-- " --\\
2004 Apr.11-June 8&0.83 & $I$ & -- " -- &-- " -- \\
2005 June 07    & 0.81  &Sloan $i^\prime$& Magellan& Pooley et al. (2006)\\
2006 Jan.5-Apr.15&0.8   & $V$ & 1.5m Maidanak& This work\\
2006 Jan.5-June 2&0.83  & $R$ & -- " -- & -- " -- \\
2006 Jan.5-Apr.15&0.85  & $I$ & -- " -- & -- " -- \\
\hline
\end{tabular}
\end{table*}

Thanks to the spatially resolved photometry of the A1 and A2 image pair in
filters $V$, $R$ and $I$, our data have made it possible to measure flux 
ratios for these components for five seasons of observations, and to study 
their behavior in time and in wavelength. As is noted above, deviations of 
flux ratios in quasar macroimages from the theoretical predictions (flux 
ratio anomalies) are presently believed to be diagnostic for detection of 
substructures in lensing galaxies, which may represent the dark matter.

\section{The A2/A1 flux ratios}
The idea to detect substructures in lensing galaxies using the anomalies of 
flux ratios is based on fundamental relationships between coordinates and 
magnifications of the quasar images, which result from the general lens 
equation (Schneider, Ehler \& Falco, 1992). These relationships have been
obtained for the first time by Schneider and Weiss (1992) and Mao (1992) 
for several "smooth" distributions of lensing potential. In principle, the 
lens equation is capable of providing six independent relationships between 
the coordinates and magnifications for a quadruple lens, but only one of 
them can be checked with the data of observations. This is the well-known 
magnification sum rule for a source within a macrocaustic cusp, when three 
close images emerge: magnification of the central image must be equal to 
the sum of magnifications of two outer images (Schneider \& Weiss 1992). 
When the source lies near a caustic fold, two images of the same brightness 
must arise (Keeton, Gaudi and Petters 2005, KGP hereafter).
\begin{figure}
\resizebox{\hsize}{!}{\includegraphics{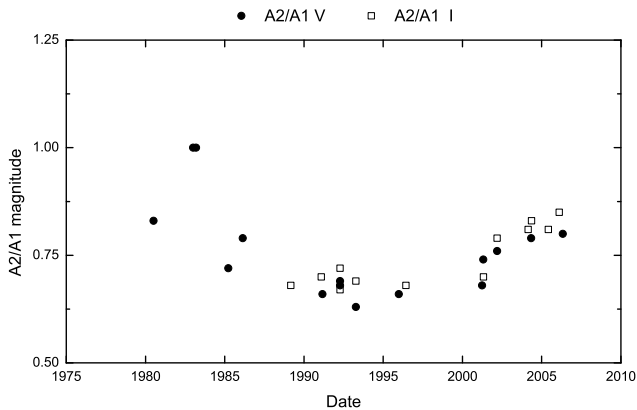}}
 \caption{A history of the A2/A1 flux ratios in PG 1115+080 from
 the data in filters $V$ and $I$ as listed in Table 1.}
\end{figure}
Since the absolute values of magnifications in macroimages are unknown
(the unlensed quasar cannot  be observed), Mao \& Schneider (1998) proposed
to use the dimensionless quantities
\begin{equation}
R_{cusp}={{|\mu_1|-|\mu_2|+|\mu_3|}\over{|\mu_1|+|\mu_2|+|\mu_3|}}=
{{F_1-F_2+F_3}\over{F_1+F_2+F_3}}
\end{equation}
for three images emerging when the source is in a caustic cusp, and
\begin{equation}
R_{fold}={{|\mu_m|-|\mu_s|}\over{|\mu_m|+|\mu_s|}}={{F_m -F_s}\over
{F_m +F_s}}
\end{equation}
for the case when the source is at the caustic fold. Indices $m$ and $s$
in the second expression denote the images at the minimum and saddle points
of the Fermat surface.

Ideally, cusp relation $R_{cusp}=0$ and fold relation $R_{fold}=0$ hold only
when the source lies exactly at the caustic cusp or fold, respectively.
In real lenses these relations hold only approximately. KGP(2003, 2005)
fulfilled a detailed study of asymptotic behaviours of the cusp and fold 
relations and calculated probability distributions of $R_{fold}$ values 
for several smooth lens models. The value of deviation of  $R_{cusp}$ and 
$R_{fold}$ from zero can be regarded as a measure of probability for the 
lensing potential to have substrutures on scales smaller than the separation 
between the closest images (KGP 2003, 2005). KGP (2005) warn, however, that 
for fold lenses, the observed violation of the fold relation may just 
mean that the source is far enough from a caustic fold.

It should be noted that, in principle, the observed anomalies of brightness
ratios in images of gravitationally lensed quasars can be explained by
other factors, such as microlensing by compact bodies and the effects of
propagation phenomena in the interstellar medium (extinction and scattering, 
scintillations). These factors are studied in details by Kochanek \& Dalal 
(2004). They concluded that substructures of cold dark matter is the best 
explanation for the flux ratio anomalies in some quadruply lensed quasars. 
They reminded also that, as was stated for the first time by Mao \& Schneider 
(1998), the fluxes of highly magnified saddle images are very sensitive to 
small gravitational perturbations as compared to low-magnification images 
and, even more importantly, these perturbations bias the fluxes towards 
demagnification, as was also noted by Schechter \& Wambsganss (2002).

\subsection{Behavior of the flux ratios in time}

The A1+A2 image pair in PG 1115+080 consists of a highly magnified minimum 
point image (A1) and saddle point image (A2) situated symmetrically with 
respect to the fold caustic very close to each other. According to theoretical 
expectations (e.g. Schneider et al. 1992), the ratio of their fluxes must be 
close to 1.

There are numerous measurements of the A2/A1 brightness ratio in PG 1115+080
made at different spectral ranges and at different epochs since 1980, which
we tried to assemble in Table 1. Some of these data have been used by Pooley
et al. (2009) to analyse a long-term history of the A2/A1 variations in
the optical band and to compare with the X-ray data, (see, e.g.,
Pooley et al. 2006, 2007, 2009). Fig. 2 from the work by Pooley et al. (2009)
demonstrates changes in the A2/A1 optical flux ratio during the time period
from 1980 to 2008, and much more dramatic changes of this ratio in X-rays.
Pooley et al. (2009) noted that, according to all the observations since the
system discovery, the A2/A1 flux ratio varied within 0.65-0.85. They did not
specify the optical spectral bands for the data in their Fig.2, however, but
it looks like they are for filter $V$.
\begin{figure*}
\centering
\resizebox{0.7\hsize}{!}{\includegraphics{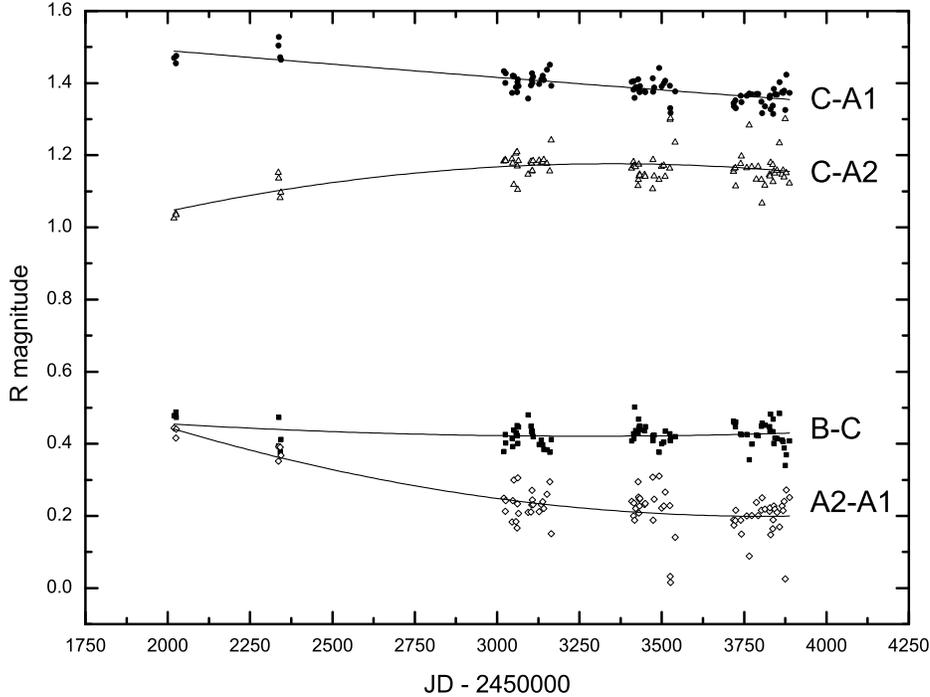}}
 \caption{Behaviors of the C-A1, C-A2, B-C and A2-A1 magnitude differences
  in time from the results of our photometry in filter $R$; approximation by 
  the second-order polynomials is shown.}
\end{figure*}

In our Table 1, the telescopes and filter bands are indicated for all estimates 
of the A2/A1 flux ratios. The table does not contain the results of the 
Chandra X-rays observations, which exhibited strong flux ratio anomaly and
can be found, e.g., in Pooley et al. (2006, 2007, 2008). The data in Table
1 are also supplemented by the estimates of the A2/A1 flux ratios obtained from 
our photometry (Tables 5-7). Flux ratios in filters $V$ and $I$ from Table
1 are displayed in Fig. 3. These flux ratios behave similarly in time for
both filters, and in general features resemble fig. 2 from Pooley et al.
(2008). Based upon their fig. 2, Pooley et al. (2008) argue that the optical
flux ratio anomaly in PG 1115+080 is slight and "nearly constant in time".
Our analysis described below have shown, however, that it is not quite so.
Our Fig. 3, where the available previous data in $V$ and $I$ are supplemented
by our measurements, shows that variations of the A2/A1 flux ratio in time
are indeed rather small and slow. However, even if we exclude a marginal
value for the date 1983, March 8 (Vanderriest et al. 1986), which equals 1
with the uncertainty of 0.1, we will have the A2/A1 flux ratio varying in
some regular manner with the amplitude of about 0.15 during the last 25 years. 
Somewhere between 1991 and 1996, the ratio reached its minimal value of about 
0.65 in filter $V$, and increased up to 0.8 by 2006. It should be noted that 
the fact that A2/A1 flux ratio varies in time is in itself an argument in 
favour of microlensing as the main reason for the anomalous flux ratio in 
PG1115+080. Also, it should be mentioned that the A2/A1 flux ratio is 
slightly but steadily higher in filter $I$ than in $V$.

 To determine which component (or components) exactly underwent microlensing,
 we addressed only our data as more homogeneous ones, and analysed behaviors
 in time of the long-term constituents of the A2-A1, C-A1, B-C and C-A2
 magnitude differences for filters $R$ and $I$. These difference light curves
 in filter $R$ are shown in Fig. 4. We did not correct the individual light 
 curves for the time delays, which are small as compared to the characteristic 
 time-scale of quasar flux variations. This might result only in some increase 
 of the data points scatter with respect to the approximating curves, which are 
 the second-order polynomials in Fig.4.
 
 The largest decrease of the magnitude
 difference is for the A2-A1 image pair - about 0.23 mag during 2001-2005.
 Pair C-A1 shows an almost linear decrease of the magnitude
 difference in time, with only 0.12 mag during 2001-2005. Since the mutual 
 brightness of images B and C was almost invariable in 2001-2006, one might 
 conclude that it is an image A1 that became fainter during this time period. 
 But the C-A2 magnitude  difference curve shows however, that, in addition 
 to the obvious dimming of image A1, brightening of image A2 makes a certain 
 contribution to the decrease of the A2-A1 magnitude difference in 2001-2005.

\begin{table}
 \centering
  \caption{Flux ratios in PG 1115+080 as predicted by the most recent
  lens models and determined from the results of our photometry in
  2006 (filter $I$); the uncertainty of our flux ratio estimates is
  0.02 for all ratios.}
  \begin{tabular}{lcccc}
  \hline
Lens model    & A2/A1 & B/A1 & C/A1 & B/C\\
\hline
Chiba (2002)  & 0.92  & 0.22 & 0.28 & 0.8\\
Chiba (2005)  & 0.92  & 0.22 & 0.28 & 0.79\\
Pooley(2006)  & 0.96  &      & 0.26 & 0.67\\
Pooley(2007)  & 0.92  & 0.21 & 0.27 & 0.78\\
\hline
This work     & 0.85  & 0.19 &0.29  & 0.68\\
\hline
\end{tabular}
\end{table}
Therefore, we can conclude that a decay of A1 and brightening of A2 took place
simultaneously in PG 1115+080 during 2001-2006. We may also conclude that it is
the A1 image that underwent microlensing in the previous years, with the maximum
near 1992-1995, as seen from Fig. 2, and the final phase in 2006 or, perhaps,
later. With the previous data taken into account (see Fig. 3 and Table 1), the
total time-scale of the 0.3-magnitude event is about 25 years. Image A2
underwent microlensing as well, with its rising branch occurring in 2001-2005.
The brightening of image A2 reached about 0.14 mag during this time period,
while the total brightening in the whole event may be larger. In calculation
of the time delays, more subtle variations of the magnitude differences during 
2004-2006 were found, (see Sec. 4.1).

It should be noted that our results are well consistent with measurements of 
the A1-A2 magnitude difference presented in Morgan et al. (2008), who reported 
approximately 0.2-mag growth of this quantity during 2001-2006. However, they 
do not present the magnitude differences between other images and A1 or A2 
separately, which has led us to a conclusion about the final phase of 
microlensing in image A1 and, seemingly, the initial phase of a microlensing 
event in image A2. This conclusion is also indirectly confirmed by the results 
of Pooley et al. (2009), who reported a dramatic rise in the X-ray flux from 
image A2 between 2001 and 2008. Larger microlensing amplitudes at shorter 
wavelengths are often detected for many lensed quasars and are known to be 
naturally explained by smaller effective sizes of quasars at shorter 
wavelengths.

The observed time-scales and amplitudes of the microlensing brightness
fluctuations are known to depend on the relative velocity of a quasar and
lensing galaxy, and on the relationship between the source size and the 
Einstein ring radius of a microlens. For PG1115+080, the expected duration of 
a microlensing event is estimated to be of the order of 10 to 20 years for 
the subsolar mass microlens (Chiba 2005), well consistent with 
that in image A1 observed in 1980-2006.

Thus, if our interpretation of the observed brightness variations in the A1, 
A2, B and C images is valid, then the A2/A1 flux ratio would be expected to
approach its undisturbed value in a few years, unless a new event takes place
in at least one image. Our estimates of the A2/A1, B/A1, C/A1 and B/C flux
ratios calculated from the photometry data of 2006, are  presented in Table 2
along with the model predictions by Chiba (2002),  Chiba et al. (2005),
Pooley  et al. (2006, 2007). The ratios for filter $I$ are presented, where
the effect  of microlensing is expected to be minimal as compared to $V$
and $R$. As is seen from Table 2, the A1/A2 flux ratio is still less than
predicted by the most recent lens models. However, when expressed in terms 
of $R_{fold}$, it would equal 0.08, which means that, according to simulations
of KGP (2005), this flux ratio is within a region admissible by a smooth
lensing potential model for the finite source distances from the caustic fold, 
that is, it is not anomalous in the sense implied by Mao\&Schneider (1998).

\subsection{Color Changes in PG1115+080}

We have also made use of our multicolor observations to analyze behaviors 
of the $V-I$ color indices of image components, which were shown to be 
indicative of the microlensing nature of brightness changes for at least the 
Q2237+0305 quasar (Vakulik et al. 2004). Since only 16 data points were 
available to build the $V-I$ versus $R$ dependency for each image component, 
we did not build them for the components separately, but combined the data 
into two sets, for A1+A2 and B+C image pairs. The resulting diagrams are 
presented in Fig. 5. In order to eliminate the magnitude difference and 
possible permanent color difference between the components in each pair, 
we shifted the data points in both plots along the $R$ axis by the values 
of the mean magnitude differences between the components.
\begin{figure}
\resizebox{\hsize}{6cm}{\includegraphics{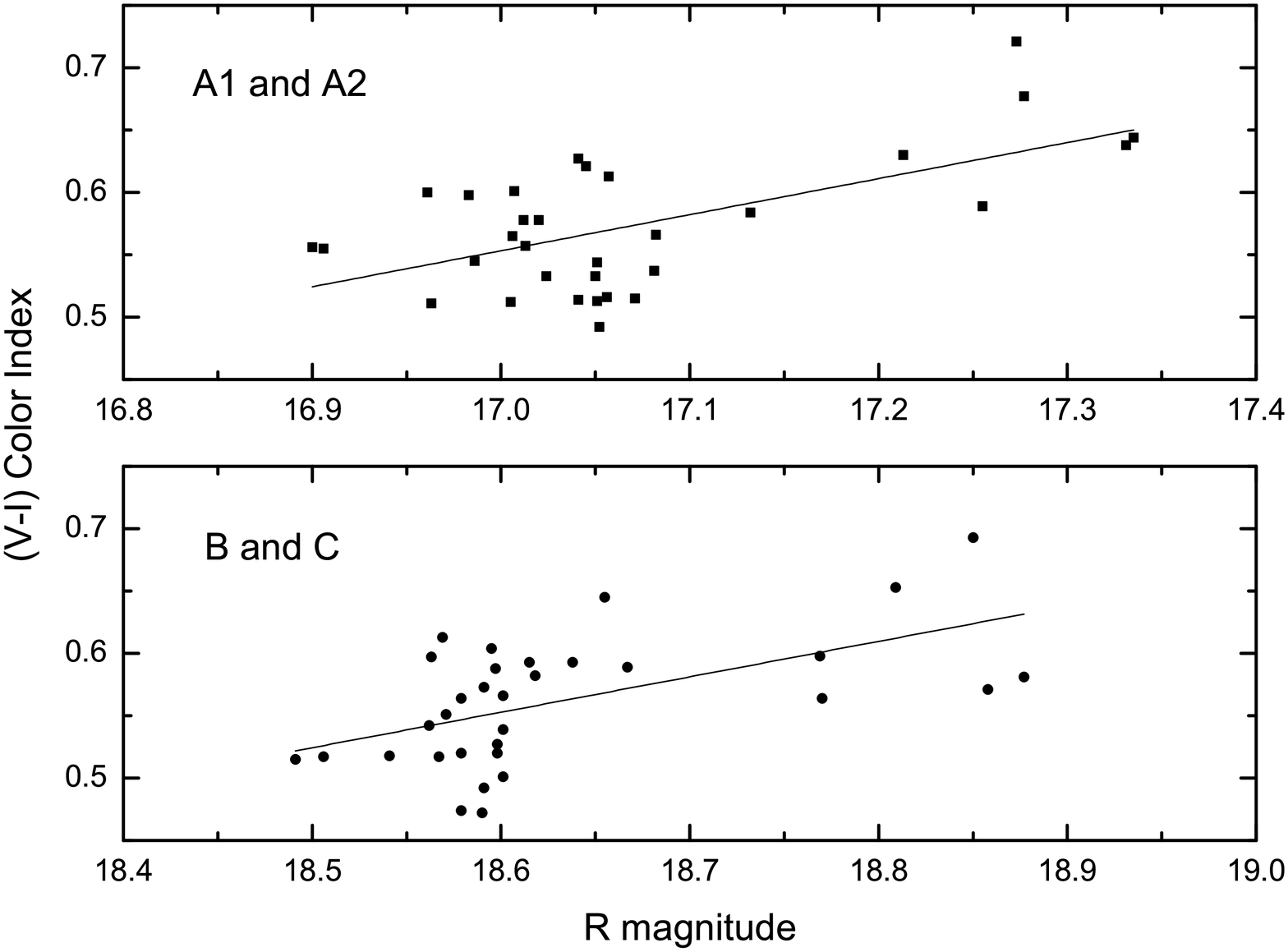}}
 \caption {$V-I$ versus $R$ diagrams for image A (upper panel) and B,C
 (bottom), illustrating a statistical dependence between the color
 indices and  magnitudes. The regression line slopes differ insignificantly
 for the A1,A2 and B,C image pairs and equal 0.29 and 0.28, with the
 correlation indices 0.61 and 0.56, respectively. }
\end{figure}
The values of $V-I$ grew with the growth of the $R$ magnitude in both 
diagrams, with the regression line slopes of 0.28-0.29. This is qualitatively 
consistent with the total 0.4-mag fading of the PG1115+080 quasar during
2001-2005: according to numerous observations, there is a common tendency 
for many quasars to become bluer at their bright phases (see, e.g. Giveon et al.
1999 and references therein). In particular, Giveon et al. (1999) presented
the $\Delta(B-R)$ versus $\Delta B$ and $\Delta(B-R)$ versus $\Delta R$
diagrams built for a subset of 21 quasars from their Palomar Green sample
consisted of 42 quasars. Their diagrams show a significant correlation
between the color and magnitude variations, with the regression line 
slopes of 0.25-0.27.

The regression line slopes for the diagrams in our Fig. 5 are also close to
that reported for Q2237+0305 in one of our previous publications (Vakulik 
et al. 2004). There is an important difference between the two quadruple 
lenses, however: the Q2237+0305 light curves are strongly dominated by 
microlensing events, while in PG1115+080, a contribution from microlensing 
activity is small as compared to the quasar intrinsic variability. We do 
not see any significant difference between the two diagrams in Fig. 5, 
though the contributions from microlensing for these image pairs are 
different. Therefore, the diagrams in Fig. 5 should be referred to 
characteristics of the PG1115+080 quasar variability rather than to 
microlensing variability, in contrast to Q2237, where they result mostly 
from microlensing.
\begin{figure}
\centering
\resizebox{\hsize}{!}
{\includegraphics{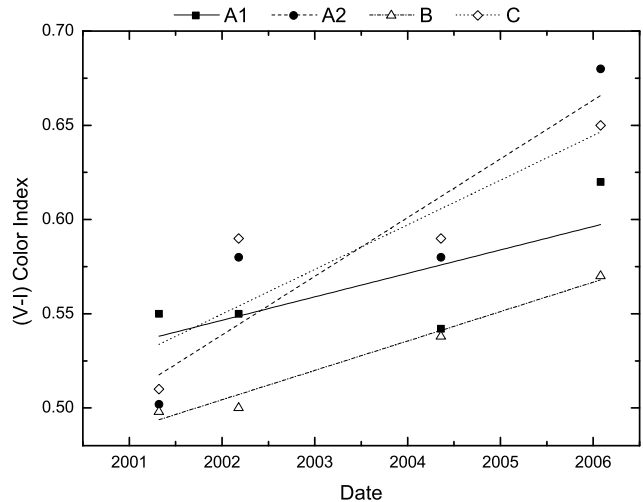}}
 \caption{A history of the long-term variations of color indices $V-I$
 in PG1115+080. Each point is a result of averaging of $V-I$ values within
 the corresponding season; the midpoints of every season are along the 
 horizontal axis. }
\end{figure}
 
 Also, we tried to analyze the long-term history of color changes in
 PG1115+080 and plotted the values of $V-I$ averaged within every season
 as functions of the corresponding time moments, which are the midpoints of 
 the seasons (Fig. 6). It should be remembered that the values of color 
 indices in this plot are accurate to only $\sim\pm 0.025$mag, therefore, 
 Fig. 6 illustrates their behaviors in time qualitatively rather than 
 quantitatively. Nevertheless, though the data points in this plot are rather 
 scattered and often overlap giving a rather intricate pattern, the general 
 features in behaviors of color indices are evident. To make them clearer, 
 we showed the corresponding linear approximations in Fig. 6. First we
 notice that the $V-I$ color indices increase in time for all the four image
 components. This conclusion seems to be valid, while the differences in 
 gradients of the $V-I$ colors with time are hardly veridical. In general, 
 Fig.6 confirms the dominating contribution from the quasar intrinsic 
 variability over microlensing activity in PG 1115+080 light curves, as was 
 indicated in the section above. Another significant conclusion from Fig.6, 
 is that the $V-I$ color index of image B remains permanently smaller that 
 those of the three other ones, while demonstrating similar increase as well. 
 We fail to explain why an image that lies the smallest distance from the 
 lensing galaxy turned out to have the bluest  color index. 
 
\section{Analysis of the new estimates of time delays in PG 1115+080}
The time delays in PG 1115+080 were determined for the first time by Schechter
et al. (1997) to be $23.7\pm 3.4$ days between B and C, and $9.4\pm 3.4$ days
between A1+A2 and C (image C is leading). Barkana (1997) re-analyzed their
data using another algorithm and reported $25^{+3.3}_{-3.8}$ days for the time
delay between B and C, and this is quite consistent  with $23.7\pm3.4$ days 
from Schechter et al. (1997). But the other time  delays, and  hence  the  
time  delay  ratio  $r_{ABC}=\tau_{AC}/\tau_{BA}$ differ significantly:  
$r_{ABC}=1.13^{+0.18}_{-0.17}$ as calculated by Barkana (1997) and $0.7\pm 0.3$ 
according to Schechter et al. (1997). Since 1997, just these values, either 
the first or the second ones, were being used to constrain the PG1115+080 model
and determine the Hubble constant.

Determination of the time delays has generated a flow of models for the system, 
(Schechter et al. 1997, Keeton \& Kochanek 1997, Courbin et al. 1997, Impey 
et al. 1998, Saha \& Wiiliams 2001, Kochanek, Keeton \& McLeod 2001, Zhao 
\& Pronk 2001, Chiba 2002, Treu \& Koopmans 2002, Yoo et al. 2005, 2006, 
Pooley et al. 2006, Miranda \& Jetzer 2007), all illustrating how strongly 
the estimated value of $H_0$ depends on the adopted mass profiles of the lens 
galaxy for the given values of time delays. 

The detailed analysis of the uncertainties in determining Hubble constant 
from the time delay lenses can be found in, e.g., Kochanek \& Schechter (2004), 
Schechter (2005), Kochanek (2002), where the paths to eliminate or at least 
to lessen the uncertainties have been also outlined. Kochanek \& Schechter 
(2004) indicated, in particular, the importance of improving the accuracy of 
time delays for PG 1115+080.
 
 The $R$ light curves taken from the 2004-2006 data (Fig. 2) clearly 
 demonstrate their  applicability to determine the time delays. As 
 compared to the data  used by Schechter et al. (1997) and Barkana (1997), 
 we were lucky to  detect the quasar brightness variation with an amplitude 
 of almost a  factor of three larger, and with rather well-sampled data 
 points within  every  season of observations. In addition, the accuracy of 
 our photometry  has made it possible to confidently detect flux variations 
 with an  amplitude as small as 0.05 mag that can be seen in the data of 2004.

 The methodology to determine the time delays in pairs is simple enough and 
 obvious. A common feature of all known methods of time delay measurements is 
 the  use, in one way or another, of the cross-correlation maximum or mutual
 dispersion minimum criteria, while they may differ in the algorithms of
 the initial data interpolation. 
\begin{table}
 \centering
 \begin{minipage}{80mm}
  \caption{The time delays (days) for PG1115+080 as reported earlier
  and in our previous work, (Vakulik et al. 2009).}
  \begin{tabular}{lccc}
  \hline
Author &  $\tau_{BA}$  &  $\tau_{AC}$  &  $\tau_{BC}$  \\
\hline
Schechter (1997)& $14.3\pm3.4$ & $9.4\pm3.4$ & $23.7\pm3.4$ \\
& & & \\
Barkana (1997)&$11.7^{+2.9}_{-3.2}$&$13.3^{+2.0}_{-2.2}$&$25.0^{+3.3}_{-3.8}$\\
& & & \\
Vakulik (2009)&$4.4^{+3.2}_{-2.5}$ &$12.0^{+2.5}_{-2.0}$&$16.4^{+3.5}_{-2.5}$\\ 
\hline
\end{tabular}
\end{minipage}
\end{table}

 Analysis of the light curves of quasar images in pairs can also be 
 applied 
 when a lens consists of more than two images. To determine the time delays 
 from the light curves shown in Fig. 2, we used another approach however, 
 as described in more detail in our previous works (Vakulik et al. 2006, 2009). 
 Here we shall only remind the fundamentals of our approach.
 
 We determined the source light curve from a joint analysis of light curves 
 of all image components. The individual time delays for pairs of  images can 
 be then determined with respect to this model source light curve jointly 
 from the corresponding system of equations. 

Since, according to predictions of all lens models and to measurements in 
X-rays by Dai et al. (2001) and Chartas et al. (2004), the time delay 
between images A1 and A2 does not exceed a small fraction of the day, 
their fluxes  were summed to form a single curve, which we call the A 
light curve. Thus, we may write the following functional for three light 
curves:
\begin{equation}
 \Phi(\Delta t, \tau_0, \tau_1, \tau_2)={1\over {3N}}\sum_{j=0}^2\sum_{i=0}
 ^N{{[m_j(t_i)+dm_j-
 f(t_i,\Delta t, \tau_j)]^2}\over{\sigma^2_j(t_i)}},
\end{equation}
where $m_j(t_i)$ are the data points in the light curve of the $j$th image 
at the time moments $t_i$; $dm_j$ and $\tau_j$ are the shifts of a 
corresponding light curve in stellar magnitude and in time, respectively, 
$N$ is a number of points in the light curves, $\Delta t$ is the parameter 
of the approximating function $f(t_i, \Delta t, \tau_j)$, and 
$\sigma^2_j(t_i)$ are the photometry errors. 
\begin{table}
 \centering
 \begin{minipage}{80mm}
  \caption{Time delay ratios $\tau_{AC}/\tau_{BA}$ and $\tau_{AC}/\tau_{BC}$  
  for PG1115+080 as predicted by several lens models (the upper part of the 
  table) and determined from the existing measurements of the time delays 
  for the system (the last three lines).}
  \begin{tabular}{lcc}
  \hline
Author &  $\tau_{AC}/\tau_{BA}$  &  $\tau_{AC}/\tau_{BC}$   \\
\hline
Schechter et al. (1997)& 1.33-1.80 & 0.57-0.64  \\
Keeton\&Kochanek (1997)& 1.35-1.47 & --         \\
Impey et al. (1998)    & 1.3       & --         \\
Chartas et al. (2004)  & 1.3       &0.56        \\
Keeton et al. (2009)   & 1.54      & 0.61       \\
\hline
Schechter et al. (1997)& 0.66      &0.40        \\
Barkana (1997)         & 1.14      & 0.53       \\
Vakulik et al. (2009)  & 2.73      & 0.73       \\
  \hline
\end{tabular}
\end{minipage}
\end{table}

We adopted $dm_0=0$ and $\tau_0=0$ in our calculations, that is, we fitted
the light curves of B and C to the A light curve, and thus, 
$dm_1$ and $dm_2$ are the magnitude differences  A--B and A--C, respectively.
At given values of $\tau_1$ and $\tau_2$, we minimize $\Phi(\Delta t, \tau_1,
\tau_2)$ in $dm_j$ and in coefficients of the approximating function.
The values of minimum of $\Phi(\Delta t, \tau_1, \tau_2)$  were being looked
for at a rectangular mesh $\tau_1, \tau_2$ with a step of 0.5 d in
preliminary calculations, and of 0.2 d at a final stage. The values of 
$\tau_1, \tau_2$ corresponding to the minimal value of $\Phi(\Delta t, \tau_1,
\tau_2)$ were adopted as the estimates of the time delays $\tau_{BA}$ and
$\tau_{AC}$.  Fig. 7 shows a distribution of $\Phi(\Delta t, \tau_1, \tau_2)-
\Phi_{min}$ in the space of parameters $\tau_{BA}$ and $\tau_{AC}$ calculated 
for parameter $\Delta t=0.12$ years. Thus, our estimates of the time delays 
that can be read out at the $\tau_{AC}$ and $\tau_{BC}$ axes against the centre 
of contours in  Fig.7, are $\tau_{BA}=4.4$, $\tau_{AC}=12.0$ days. The time 
delay $\tau_{BC}$ is not an independent quantity in our method, and can be 
determined as a linear combination $\tau_{BC}=\tau_{BA}+\tau_{AC}$, that is, 
$\tau_{BC}=16.4$ days. 

 To test our method for robustness and absence of systematics, and to estimate 
 the accuracy inherent in our time delay measurements, we fulfilled a numerical 
 simulation as described in detail in (Vakulik et al. 2009). The simulated 
 light curves of the components were obtained by shifting the approximating 
 curve $f(t_i, \Delta t, \tau_j)$ by the proper time delays $\tau_1, \tau_2$ 
 and magnitude differences, and by adding random quantities to imitate the 
 photometry errors. We simulated 2000 light curves synthesized as described 
 above, and calculated the resulting time delays using the procedure, which 
 was exactly the same as in the  analysis of the actual light curves. The 
 results of simulations were used to build the distribution functions for 
 errors and to estimate the 95-percent confidence intervals.
 
 The final values of the time delays and the corresponding uncertainties are 
presented in Table 3, where they can be compared with the estimates reported 
by Schechter et al. (1997) and  Barkana (1997).

\begin{table*}
 \centering
  \caption{Time delays as predicted by the lens models calculated by 
  Schechter et al.(1997) and the values of the Hubble constant $H_0$ 
  obtained from comparison of these time delays with those obtained by 
  Schechter et al. (1997), columns  2-4; the $H_0$ values calculated 
  for the same lens models with the time delays determined in this work 
  (columns 5-7).}
  \begin{tabular}{ccccccc}
  \hline
\multicolumn{4}{c}{Lens models, $\tau$(days) and $H_0$(Schechter et al.1997
)}&
\multicolumn{3}{c}{$H_0$ with $\tau_{AC}$ and  $\tau_{BC}$ from this work}\\
\hline
Model & $\tau_{AC}(days)$ & $\tau_{BC}(days)$&$H_0$(km s$^{-1}$Mpc$^{-1}$)&
 $H_0$
($\tau_{AC}=12.0^d$) & $H_0$($\tau_{BC}=16.4^d$) & $H_0$(mean)\\
\hline
PMXS & 12.5&19.9 & 84 &104 & 121 & 113\\
ISXS & 6.6 &10.4 & 44 & 55 & 63  & 59\\
ISEP & 9.7 &15.1 & 64 & 81 & 92  & 86\\
ISIS & 5.6 & 9.7 & 41 & 47 & 59  & 53\\
ISIS+& 5.7 & 10  & 42 & 48 & 61  & 54\\
\hline
\end{tabular}
\end{table*}

\begin{figure}
\resizebox{0.85\hsize}{!}{\includegraphics{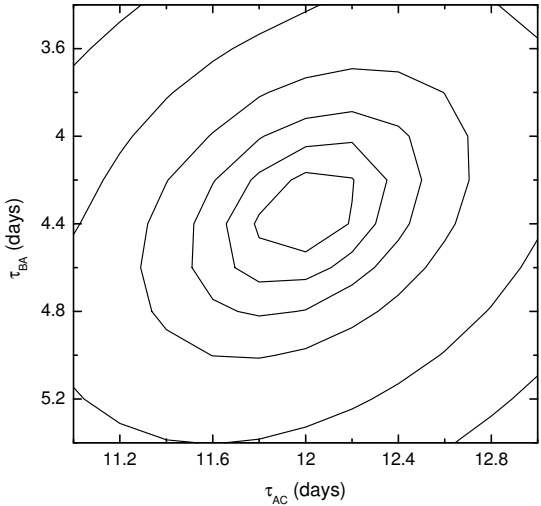}}
 \caption{Distribution of $\Phi(\Delta t, \tau_1, \tau_2)-\Phi_{min}$ in the
 space of parameters $\tau_{AC}$ and $\tau_{AB}$. The innermost contour 
 corresponds to $\Phi(\Delta t, \tau_1, \tau_2)-\Phi_{min}$ equalling 0.0001, 
 with every next level twice as much than preceding.}
\end{figure}

 It is interesting to note that using only the data of 2004, where a 
 small-amplitude turn-over in the light curves is detected, we obtained
 $\tau_{BA}=5.0$ days, $\tau_{AC}=9.4$ days, and $\tau_{BC}=14.4$days,
 consistent with the estimates obtained from the whole data set. However, 
 simulation of errors for only the data of 2004 demonstrates noticeably larger 
 uncertainties, as compared to those calculated from the entire light curve.

 The light curves of images A, B and C shifted by the corresponding 
 time delays and reduced to image A in magnitude are shown in figure 2 from 
 our previous paper (Vakulik et al 2009) for the approximating function 
 parameter $\Delta t=0.12$ years. As is seen from this picture, the data 
 points for all the three images are very well consistent with each other 
 and with the approximating curve. 

Thus we obtained the time delay values, which differ noticeably from those
reported by Schechter et al. (1997) and Barkana (1997) and used in a variety
of models of many authors to derive the Hubble constant value. The largest 
differences are for $\tau_{BC}$ and $\tau_{BA}$: our estimate of $\tau_{BC}$ 
is a factor of 1.5 smaller, while for $\tau_{BA}$, it is almost three times 
smaller as compared to the results of Schechter (1997) and Barkana (1997). 
Meanwhile, our values of $\tau_{AC}$ are rather similar to those of Schechter 
and, especially, of Barkana. 

As is noted in Introduction, the time delay between one of the image pairs, 
say, $\tau_{BC}$, can be used to determine $H_0$, while the time delay 
ratio $r_{ABC}=\tau_{AC}/\tau_{BA}$ is independent of the $H_0$ value and 
can be used to constrain the lens model. Most of the PG1115+080 macromodels 
are consistent in predicting $r_{ABC}$ to within 0.15. As far as can be 
expected from Table 3, the three measurements of time delays do not provide 
the time delay ratios consistent with each other and with model predictions.
This can be seen from Table 4, where we collected several model predictions 
for $r_{ABC}$  together with the measurements presented by Schechter et al.
(1997), Barkana (1997) and Vakulik et al. (2009). Also, we presented here the 
time delay ratios $r_{CBA}=\tau_{AC}/\tau_{BC}$, which  are connected with 
$r_{ABC}$ by a relationship $r_{CBA}=r_{ABC}/(1+r_{ABC})$. These quantities 
demonstrate better agreement between model predictions and measurements. 

The time delay ratios calculated from measurements of Schechter and Barkana 
are seen to be lower as compared to the model predictions, while our 
measurements provide the estimates of this quantities exceeding the model 
predictions, especially for $r_{ABC}=\tau_{AC}/\tau_{BA}$, which is as 
large as 2.73 from our data. One should admit that such discrepancy is too 
large, since the largest  $r_{ABC}$ we have found in the literature is that 
calculated by Schechter et al. (1997) for their isothermal ellipsoid model - 
$r_{ABC}=1.8$. 

The reason for this becomes qualitatively clear when addressing the data in 
Table 3: the shortest time delay $\tau_{BA}$ is measured with almost the same 
absolute error as the longest one, $\tau_{BC}$, that is, of all the three time 
delays, $\tau_{BA}$ has the largest relative error. Therefore, we are far from 
arguing our value of $\tau_{BA}$ to be more trustworthy that those obtained by 
Schechter and Barkana. We regard, however, that the values of $\tau_{BC}$ and 
$\tau_{AC}$ are more reliable and trustworthy. Also, it should be noted that 
none of the macrolens models we could found in the literature predicts the 
values of $\tau_{BC}$ larger than 18d, - the value 19.9d from the unrealistic 
point-mass model in Schechter et al. (1997) may hardly be taken into account.

In the framework of the present publication, we did not intend to either 
propose an extra exotic lens model or recalculate the most popular ones to 
derive the new estimate of the Hubble constant, but instead, we have made 
use of the results of Schechter et al. (1997), who calculated five models 
for the gravitational potential of PG1115+080. The time delays $\tau_{AC}$ 
and $\tau_{BC}$ as predicted by their five models for $\Omega=1$ and 
$H_0=100$km s$^{-1}$Mpc$^{-1}$, and the corresponding estimates of $H_0$ 
obtained with their time delays are shown in Table 5 (columns 2-4).
We remind that, according to Schechter et al. (1997), model PMXS means a 
point mass with external shear, the ISXS model is an isothermal sphere 
with external shear, and ISEP is an isothermal elliptical potential. The 
ISIS model uses a second isothermal sphere for a group of galaxies at 
approximately the same redshift as the main lensing galaxy (Young et al. 
1981; Henry \& Heasley 1986) to represent shear. In the ISIS+ model, the 
uncertainty in the galaxy position is not regarded to be negligible, as 
in the ISXS and ISIS, but its coordinates were taken as two additional 
free parameters.

Columns 5 to 7 of Table 5 contain the $H_0$ values estimated for the
Schechter et al. models with our values of the time delays for image pairs
AC and BC separately (column 5 and 6), and the average between the pairs
(column 7). We may conclude that, as could be expected, our new estimates 
of $\tau_{BC}$ and $\tau_{AC}$ provide higher values of Hubble constant, 
which are closer to the most recent value obtained in the HST Key Project 
from observations of Cepheids (Freedman et al. 2001).

\section {Discussion and conclusions}

From our new data of excellent quality, we had a possibility 
to discriminate between the quasar intrinsic and microlensing brightness 
and color variations for PG 1115+080, and to obtain new estimates of the 
time delays. It may immediately be seen from our $R$ light curves, especially
for 2004-6 seasons, that significant brightness fluctuations with amplitudes 
exceeding the typical error bars of the data points, have been detected. 
This allows us to recompute the values of $H_0$ calculated with the 
previous estimates of the time delays for this gravitational lens system. 

In particular, we find the following:
\begin{itemize}
 \item We have studied behaviors of brightness ratios of all the components
 both in time and in wavelength. We report microlensing in A1 with an
 amplitude of about 0.3 mag in filter $R$ on a time-scale of 25 years. The
 magnification peak in A1 took place in 1992-1995, with the subsequent fading in
 2001-2006. The image A1 flux may be expected to reach its undisturbed value
 by 2006 or later. A microlensing event was apparently observed in image A2 as
 well, with its rising branch in 2001-2005, when image A2 brightened by
 approximately 0.15 mag. The time scales and amplitudes of both events are
 consistent with those predicted for this object for the Solar mass microlenses
 (Schechter 2004). We also notice that a similar microlensing amplitude was 
 detected  for another quasar with similar redshift distances to the quasar 
 and lens, -  for  the First Lens Q0957+561 (Schild \& Smith 1991). In fitting 
 the 2004-2006 data points to the approximating  function, very subtle signs 
 of microlensing have also been found in image B.
 
 \item Therefore, deviation of the observed A2/A1 flux ratio from that 
 predicted by  most of the lens models can be well explained by microlensing 
 events. An  additional  contribution to the flux ratio "anomaly" may be 
 expected from  the source  position  with respect to the caustic fold: when 
 expressed in terms of  $R_{fold}$ (Eq. 2), the brightness difference between 
 A1 and A2 would equal 0.08, which means that, according to simulations of 
 KGP (2005), it is within a region admissible by a smooth lensing potential
 model, that is, it is not  anomalous in the sense implied first by Mao \& 
 Schneider (1998).

\item We have made use of observations in other filters available for some
dates to analyse behaviors of color indices of the images. The $V-I$ versus
$R$ diagrams built for pairs A1+A2 and B+C demonstrate the known tendency of
quasars to become bluer at their bright phases, with no signs of any
contribution from microlensing: the diagrams for both image pairs are  
nearly identical. This can be explained either by poor statistics (the data 
in all the three filters are available for only 16 nights, providing rather 
poor correlation as indicated in Fig. 5 caption), or by small amplitudes of 
the microlensing events under consideration, or by both reasons. 

\item An interesting feature of the behaviours of color indices should be
noted. While all images demonstrate growth of their color indices in time,  
the $V-I$ color indices of image B are slightly but steadily less than 
those of other images. This is rather unexpected, since image B is located 
the closest distance to the main lensing galaxy.

\item The time delays for PG1115+080 obtained from our monitoring
data in 2004-2006 differ from those determined by Schechter et al. (1997) and
Barkana (1997) earlier. The differences for $\tau_{BA}$ and $\tau_{BC}$ are
well beyond the uncertainties reported in both publications and determined
in the present work. While our time delay estimates for images A and C are
rather close to the two previous ones, the delays for two other image pairs
can not be regarded as consistent even marginally.

\item As could be expected, our estimates of time delays $\tau_{AC}$ and 
$\tau_{BC}$ result in larger $H_0$ values than those reported by Schechter 
et al. (1997) with their estimates of time delays and with the ISXS, ISIS and
ISIS+ models by Schechter et al. (1997). The new estimates of $H_0$ are more 
consistent with the most recent $H_0$ value obtained in the HST Key project 
(Freedman et al. 2001).

\item The new estimates of time delays in PG1115+080 provide additional 
support for the family of models close to isothermal. As analyzed in details 
by Kochanek and Schechter (2004), the estimates of $H_0$ with the use of 
time delay lenses are bounded by two limiting models: models with less 
dark matter (more centrally concentrated mass profiles) produce higher 
values of $H_0$ than those with more dark matter. In particular, the 
constant mass-to-light ratio models set an upper limit on estimates of $H_0$,
while the isothermal mass distribution models are responsible for the lower
limit of $H_0$. Our result is very important in this respect, since an
isothermal model is preferred for the lensing galaxy in PG1115+080 for the
reasons listed by Schechter (2005): (1)the velocity  dispersions observed
for an ensemble of lensing galaxies are consistent with the fundamental plane
relations for ellipticals; (2) a majority  of the nearby galaxies, as well
as those lensing galaxies for which the radial  mass distributions can be
measured, are very nearly  isothermal. Since the PG1115+080 lensing galaxy
is by no means unusual, the isothermal hypothesis is most probable. 
\end{itemize}

In conclusion, recently published (Morgan et al. 2008) observations of
PG 115+080 in filter $R$ during almost exactly the same time periods in
2004-2006 should be mentioned. We have used their table 3 photometry
to compare to our light curves. The quasar brightness fluctuations
which allowed us to determine the time delays are seen in their A1+A2 
light curve quite well, but become undetectable in the B and C light 
curves because of a much larger scatter of the data points.
   
\section*{Acknowledgements}
The authors from Ukraine and Uzbekistan are grateful to the Science and
Technology Center in Ukraine (STCU grant U127) which made possible
observations at the Maidanak Observatory and the further processing of the 
data. The authors from Ukraine thank also the National Program 
"CosmoMicroPhysics", and the authors from Uzbekistan are thankful to the 
DFG grant 436 Usb 113/5/0-1. This work has been also supported by the 
Russian Foundation for Basic Research (RFBR) grant No.09-02-00244 and 
No.06-02-16857.

Also, the authors would like to express their sincere gratitude to the 
anonymous referee for very constructive and helpful comments.
\begin{table*}
 \centering

  \caption{Photometry of the PG1115+030 images in filter $V$. }
  \begin{tabular}{@{}llcllll@{}}
  \hline\noalign{\smallskip}
  Date  & JD& Seeing (arcsec)& A1 & A2 & B & C \\
  \hline\noalign{\smallskip}
20-04-2001&2452019&1.47 &17.025$\pm$0.097&17.025$\pm$0.079&18.8$\pm$0.09   &18.232$\pm$0.088\\
26-04-2001&2452025&1.258&16.868$\pm$0.021&17.19$\pm$0.017 &18.736$\pm$0.02 &18.296$\pm$0.019\\
27-04-2001&2452026&1.287&16.874$\pm$0.037&17.214$\pm$0.03 &18.781$\pm$0.035&18.298$\pm$0.034\\
03-03-2002&2452336&1.171&16.889$\pm$0.024&17.252$\pm$0.019&18.782$\pm$0.022&18.406$\pm$0.022\\
08-03-2002&2452341&1.198&16.985$\pm$0.046&17.215$\pm$0.038&18.819$\pm$0.044&18.435$\pm$0.043\\
17-01-2004&2453021&1.038&16.972$\pm$0.007&17.185$\pm$0.006&18.782$\pm$0.007&18.403$\pm$0.007\\
13-02-2004&2453048&1.236&17.0  $\pm$0.02 &17.248$\pm$0.017&18.815$\pm$0.019&18.424$\pm$0.018\\
22-02-2004&2453057&1.262&16.973$\pm$0.026&17.28$\pm$0.022 &18.843$\pm$0.024&18.412$\pm$0.024\\
26-02-2004&2453061&1.029&17.04$\pm$0.014 &17.221$\pm$0.012&18.828$\pm$0.013&18.418$\pm$0.013\\
27-02-2004&2453062&0.845&17.005$\pm$0.004&17.243$\pm$0.003&18.831$\pm$0.004&18.404$\pm$0.004\\
28-02-2004&2453063&1.15 &17.021$\pm$0.019&17.224$\pm$0.016&18.856$\pm$0.018&18.41$\pm$0.018\\
01-03-2004&2453065&1.059&17.008$\pm$0.014&17.237$\pm$0.012&18.832$\pm$0.013&18.404$\pm$0.013\\
10-04-2004&2453105&1.02 &16.934$\pm$0.006&17.187$\pm$0.005&18.775$\pm$0.006&18.353$\pm$0.006\\
11-04-2004&2453106&1.183&16.939$\pm$0.013&17.187$\pm$0.011&18.792$\pm$0.012&18.356$\pm$0.012\\
14-04-2004&2453109&0.947&16.947$\pm$0.005&17.168$\pm$0.004&18.754$\pm$0.005&18.352$\pm$0.005\\
03-05-2004&2453128&1.203&16.932$\pm$0.03 &17.196$\pm$0.024&18.699$\pm$0.028&18.38$\pm$0.028\\
09-05-2004&2453134&0.842&16.96$\pm$0.004 &17.214$\pm$0.003&18.751$\pm$0.004&18.405$\pm$0.003\\
13-05-2004&2453138&0.842&16.972$\pm$0.005&17.218$\pm$0.004&18.772$\pm$0.005&18.401$\pm$0.005\\
17-05-2004&2453142&1.322&16.942$\pm$0.026&17.262$\pm$0.022&18.829$\pm$0.025&18.437$\pm$0.024\\
26-05-2004&2453151&1.169&16.95$\pm$0.023 &17.279$\pm$0.019&18.792$\pm$0.022&18.414$\pm$0.022\\
04-06-2004&2453160&1.347&17.077$\pm$0.027&17.172$\pm$0.022&18.806$\pm$0.025&18.438$\pm$0.025\\
08-06-2004&2453164&1.233&17.11$\pm$0.032 &17.145$\pm$0.027&18.845$\pm$0.03 &18.482$\pm$0.029\\
05-01-2006&2453740&1.038&17.316$\pm$0.008&17.561$\pm$0.006&19.09$\pm$0.007 &18.72$\pm$0.007\\
08-03-2006&2453802&1.093&17.313$\pm$0.011&17.526$\pm$0.009&19.083$\pm$0.01 &18.657$\pm$0.01\\
15-04-2006&2453840&0.979&17.2  $\pm$0.01 &17.453$\pm$0.008&18.977$\pm$0.009&18.576$\pm$0.009\\
\hline
\end{tabular}
\end{table*}
\begin{table*}
 \centering

  \caption{Photometry of the PG1115+030 images in filter $I$. }
  \begin{tabular}{@{}llcllll@{}}
  \hline\noalign{\smallskip}
  Date  & JD& Seeing (arcsec)& A1 & A2 & B & C \\
  \hline\noalign{\smallskip}
26-04-2001&2452025&1.197&16.313$\pm$0.026&16.698$\pm$0.021&18.262$\pm$0.025&17.781$\pm$0.024\\
27-04-2001&2452026&1.144&16.318$\pm$0.018&16.699$\pm$0.015&18.261$\pm$0.018&17.781$\pm$0.017\\
03-03-2002&2452336&1.01 &16.362$\pm$0.012&16.654$\pm$0.01 &18.282$\pm$0.011&17.817$\pm$0.011\\
04-03-2002&2452337&1.351&16.421$\pm$0.07 &16.578$\pm$0.058&18.337$\pm$0.066&17.845$\pm$0.065\\
05-03-2002&2452338&0.858&16.35 $\pm$0.018&16.662$\pm$0.014&18.312$\pm$0.017&17.793$\pm$0.016\\
07-03-2002&2452340&1.288&16.371$\pm$0.037&16.695$\pm$0.03 &18.317$\pm$0.034&17.841$\pm$0.034\\
08-03-2002&2452341&1.139&16.399$\pm$0.035&16.661$\pm$0.028&18.285$\pm$0.032&17.836$\pm$0.032\\
11-03-2002&2452344&1.043&16.417$\pm$0.023&16.66 $\pm$0.019&18.329$\pm$0.021&17.853$\pm$0.021\\
14-03-2002&2452347&0.928&16.42 $\pm$0.015&16.67 $\pm$0.012&18.333$\pm$0.014&17.828$\pm$0.014\\
11-04-2004&2453106&1.136&16.427$\pm$0.01 &16.622$\pm$0.008&18.265$\pm$0.01 &17.814$\pm$0.01\\
14-04-2004&2453109&0.773&16.414$\pm$0.005&16.623$\pm$0.004&18.262$\pm$0.005&17.801$\pm$0.005\\
03-05-2004&2453128&1.125&16.418$\pm$0.031&16.598$\pm$0.026&18.182$\pm$0.029&17.767$\pm$0.029\\
09-05-2004&2453134&0.747&16.427$\pm$0.005&16.636$\pm$0.004&18.25 $\pm$0.005&17.832$\pm$0.005\\
13-05-2004&2453138&0.765&16.428$\pm$0.005&16.64 $\pm$0.005&18.252$\pm$0.005&17.835$\pm$0.005\\
17-05-2004&2453142&1.166&16.426$\pm$0.021&16.661$\pm$0.017&18.265$\pm$0.02 &17.833$\pm$0.019\\
26-05-2004&2453151&1.057&16.437$\pm$0.016&16.652$\pm$0.013&18.253$\pm$0.015&17.832$\pm$0.015\\
04-06-2004&2453160&1.222&16.464$\pm$0.054&16.606$\pm$0.045&18.213$\pm$0.051&17.845$\pm$0.05\\
08-06-2004&2453164&1.049&16.526$\pm$0.027&16.588$\pm$0.023&18.256$\pm$0.026&17.837$\pm$0.025\\
05-01-2006&2453740&1.046&16.672$\pm$0.009&16.84 $\pm$0.007&18.509$\pm$0.008&18.026$\pm$0.008\\
08-03-2006&2453802&0.972&16.675$\pm$0.008&16.849$\pm$0.007&18.512$\pm$0.008&18.004$\pm$0.008\\
15-04-2006&2453840&0.804&16.611$\pm$0.006&16.823$\pm$0.005&18.413$\pm$0.005&17.978$\pm$0.005\\
\hline
\end{tabular}
\end{table*}

\clearpage
\onecolumn
\begin{longtable}[]{@{}llcllll@{}}
\caption{Photometry of the PG 1115+030 images in filter $R$. }\\
 \hline\noalign{\smallskip}

Date   & JD  & Seeing (arcsec)& A1 & A2 & B & C \\
  \hline\noalign{\smallskip}
20-04-01&2452019&1.226&16.625$\pm$0.011&17.069$\pm$0.009&18.573$\pm$0.011&18.095$\pm$0.01\\
26-04-01&2452025&1.277&16.636$\pm$0.013&17.052$\pm$0.01 &18.579$\pm$0.012&18.091$\pm$0.012\\
27-04-01&2452026&1.126&16.63 $\pm$0.008&17.071$\pm$0.007&18.579$\pm$0.008&18.106$\pm$0.008\\
03-03-02&2452336&1.065&16.693$\pm$0.014&17.045$\pm$0.011&18.59 $\pm$0.013&18.197$\pm$0.013\\
04-03-02&2452337&1.359&16.687$\pm$0.038&17.079$\pm$0.031&18.688$\pm$0.036&18.215$\pm$0.035\\
08-03-02&2452341&1.204&16.691$\pm$0.049&17.081$\pm$0.04 &18.541$\pm$0.046&18.163$\pm$0.046\\
11-03-02&2452344&1.266&16.713$\pm$0.034&17.081$\pm$0.028&18.59 $\pm$0.032&18.178$\pm$0.031\\
17-01-04&2453021&1.27 &16.762$\pm$0.031&17.011$\pm$0.026&18.574$\pm$0.029&18.195$\pm$0.029\\
21-01-04&2453025&1.45 &16.785$\pm$0.034&16.998$\pm$0.029&18.612$\pm$0.032&18.186$\pm$0.032\\
22-01-04&2453026&0.81 &16.767$\pm$0.004&17.009$\pm$0.004&18.596$\pm$0.004&18.194$\pm$0.004\\
10-02-04&2453045&1.254&16.832$\pm$0.027&17.015$\pm$0.023&18.62 $\pm$0.026&18.205$\pm$0.025\\
13-02-04&2453048&1.174&16.799$\pm$0.019&17.041$\pm$0.015&18.611$\pm$0.018&18.219$\pm$0.017\\
16-02-04&2453051&1.451&16.787$\pm$0.036&17.087$\pm$0.03 &18.644$\pm$0.034&18.206$\pm$0.034\\
22-02-04&2453057&1.361&16.832$\pm$0.037&17.016$\pm$0.031&18.652$\pm$0.035&18.221$\pm$0.034\\
26-02-04&2453061&1.065&16.841$\pm$0.021&17.007$\pm$0.018&18.667$\pm$0.02 &18.216$\pm$0.02\\
27-02-04&2453062&0.828&16.799$\pm$0.004&17.032$\pm$0.003&18.625$\pm$0.004&18.202$\pm$0.004\\
28-02-04&2453063&1.225&16.766$\pm$0.028&17.072$\pm$0.023&18.578$\pm$0.026&18.177$\pm$0.026\\
01-03-04&2453065&1.075&16.813$\pm$0.018&17.02 $\pm$0.015&18.652$\pm$0.017&18.205$\pm$0.017\\
30-03-04&2453094&1.157&16.803$\pm$0.013&17.013$\pm$0.011&18.64 $\pm$0.012&18.16 $\pm$0.012\\
08-04-04&2453103&1.208&16.772$\pm$0.01 &16.983$\pm$0.008&18.615$\pm$0.01 &18.166$\pm$0.01\\
10-04-04&2453105&1.033&16.754$\pm$0.005&16.985$\pm$0.004&18.597$\pm$0.005&18.163$\pm$0.005\\
11-04-04&2453106&1.032&16.735$\pm$0.006&17.006$\pm$0.005&18.598$\pm$0.006&18.162$\pm$0.006\\
12-04-04&2453107&1.145&16.747$\pm$0.011&16.991$\pm$0.009&18.578$\pm$0.011&18.148$\pm$0.01\\
14-04-04&2453109&0.927&16.754$\pm$0.004&16.986$\pm$0.003&18.591$\pm$0.004&18.171$\pm$0.004\\
03-05-04&2453128&1.134&16.771$\pm$0.023&16.983$\pm$0.019&18.567$\pm$0.022&18.169$\pm$0.021\\
09-05-04&2453134&0.761&16.78 $\pm$0.003&17.012$\pm$0.003&18.601$\pm$0.003&18.191$\pm$0.003\\
13-05-04&2453138&0.827&16.781$\pm$0.004&17.02 $\pm$0.003&18.598$\pm$0.004&18.201$\pm$0.003\\
17-05-04&2453142&1.244&16.786$\pm$0.017&17.007$\pm$0.014&18.579$\pm$0.016&18.195$\pm$0.016\\
26-05-04&2453151&1.005&16.781$\pm$0.013&17.041$\pm$0.011&18.601$\pm$0.012&18.218$\pm$0.012\\
04-06-04&2453160&1.144&16.787$\pm$0.013&17.082$\pm$0.011&18.615$\pm$0.012&18.238$\pm$0.012\\
08-06-04&2453164&1.138&16.862$\pm$0.023&17.013$\pm$0.019&18.667$\pm$0.022&18.255$\pm$0.022\\
08-02-05&2453409&0.84 &17.018$\pm$0.004&17.258$\pm$0.003&18.83 $\pm$0.004&18.422$\pm$0.004\\
13-02-05&2453414&1.192&17.06 $\pm$0.011&17.26 $\pm$0.009&18.87 $\pm$0.01 &18.442$\pm$0.01\\
14-02-05&2453415&1.344&17.021$\pm$0.018&17.256$\pm$0.015&18.842$\pm$0.017&18.427$\pm$0.017\\
16-02-05&2453417&1.461&17.08 $\pm$0.049&17.268$\pm$0.041&18.941$\pm$0.047&18.439$\pm$0.046\\
20-02-05&2453421&0.888&17.042$\pm$0.007&17.263$\pm$0.005&18.868$\pm$0.006&18.431$\pm$0.006\\
27-02-05&2453428&1.011&17.027$\pm$0.01 &17.322$\pm$0.008&18.867$\pm$0.01 &18.438$\pm$0.01\\
28-02-05&2453429&1.185&17.038$\pm$0.015&17.29 $\pm$0.012&18.89 $\pm$0.014&18.422$\pm$0.014\\
01-03-05&2453430&1.232&17.05 $\pm$0.013&17.258$\pm$0.011&18.878$\pm$0.012&18.433$\pm$0.012\\
02-03-05&2453431&0.902&17.036$\pm$0.006&17.282$\pm$0.005&18.864$\pm$0.005&18.424$\pm$0.005\\
04-03-05&2453433&1.017&17.044$\pm$0.006&17.272$\pm$0.005&18.869$\pm$0.005&18.419$\pm$0.005\\
05-03-05&2453434&0.944&17.046$\pm$0.006&17.294$\pm$0.004&18.878$\pm$0.005&18.438$\pm$0.005\\
19-03-05&2453448&1.097&17.057$\pm$0.008&17.289$\pm$0.007&18.872$\pm$0.007&18.435$\pm$0.007\\
21-03-05&2453450&0.944&17.037$\pm$0.009&17.271$\pm$0.008&18.86 $\pm$0.009&18.412$\pm$0.009\\
13-04-05&2453473&1.168&17.017$\pm$0.01 &17.324$\pm$0.008&18.852$\pm$0.009&18.431$\pm$0.009\\
14-04-05&2453474&0.948&17.063$\pm$0.004&17.251$\pm$0.004&18.848$\pm$0.004&18.439$\pm$0.004\\
17-04-05&2453477&1.016&17.06 $\pm$0.011&17.306$\pm$0.009&18.872$\pm$0.01 &18.448$\pm$0.01\\
02-05-05&2453492&1.348&17.047$\pm$0.029&17.357$\pm$0.024&18.866$\pm$0.027&18.489$\pm$0.027\\
10-05-05&2453500&1.161&17.101$\pm$0.012&17.323$\pm$0.01 &18.892$\pm$0.011&18.492$\pm$0.011\\
16-05-05&2453506&1.11 &17.1  $\pm$0.016&17.328$\pm$0.013&18.904$\pm$0.014&18.5  $\pm$0.014\\
20-05-05&2453510&1.208&17.101$\pm$0.027&17.367$\pm$0.022&18.943$\pm$0.025&18.508$\pm$0.025\\
03-06-05&2453524&1.145&17.123$\pm$0.018&17.352$\pm$0.015&18.944$\pm$0.017&18.516$\pm$0.017\\
05-06-05&2453526&1.381&17.225$\pm$0.041&17.257$\pm$0.035&18.965$\pm$0.039&18.556$\pm$0.038\\
06-06-05&2453527&1.316&17.231$\pm$0.03 &17.246$\pm$0.025&18.964$\pm$0.028&18.549$\pm$0.028\\
20-06-05&2453541&1.421&17.188$\pm$0.104&17.329$\pm$0.087&18.985$\pm$0.098&18.565$\pm$0.097\\
14-12-05&2453718&0.824&17.116$\pm$0.006&17.305$\pm$0.005&18.922$\pm$0.006&18.46 $\pm$0.005\\
15-12-05&2453719&1.111&17.12 $\pm$0.008&17.294$\pm$0.007&18.915$\pm$0.008&18.456$\pm$0.008\\
19-12-05&2453723&1.181&17.111$\pm$0.018&17.298$\pm$0.015&18.911$\pm$0.016&18.463$\pm$0.016\\
20-12-05&2453724&1.072&17.09 $\pm$0.023&17.306$\pm$0.019&18.881$\pm$0.022&18.421$\pm$0.022\\
05-01-06&2453740&1.061&17.085$\pm$0.007&17.273$\pm$0.006&18.877$\pm$0.006&18.45 $\pm$0.006\\
07-01-06&2453742&1.273&17.104$\pm$0.026&17.254$\pm$0.022&18.876$\pm$0.024&18.451$\pm$0.024\\
23-01-06&2453758&1.11 &17.086$\pm$0.008&17.286$\pm$0.007&18.877$\pm$0.008&18.451$\pm$0.008\\
31-01-06&2453766&1.498&17.138$\pm$0.115&17.226$\pm$0.096&18.865$\pm$0.108&18.509$\pm$0.106\\
07-02-06&2453773&0.954&17.072$\pm$0.004&17.273$\pm$0.004&18.84 $\pm$0.004&18.441$\pm$0.004\\
21-02-06&2453787&0.948&17.058$\pm$0.004&17.295$\pm$0.003&18.852$\pm$0.004&18.428$\pm$0.004\\
27-02-06&2453793&1.407&17.069$\pm$0.018&17.27 $\pm$0.015&18.862$\pm$0.017&18.439$\pm$0.017\\
08-03-06&2453802&0.972&17.061$\pm$0.007&17.277$\pm$0.006&18.858$\pm$0.007&18.409$\pm$0.007\\
11-03-06&2453805&0.989&17.032$\pm$0.02 &17.282$\pm$0.016&18.805$\pm$0.019&18.349$\pm$0.018\\
19-03-06&2453813&0.932&17.038$\pm$0.006&17.257$\pm$0.005&18.826$\pm$0.006&18.374$\pm$0.006\\
02-04-06&2453827&0.902&17.014$\pm$0.004&17.227$\pm$0.004&18.821$\pm$0.004&18.373$\pm$0.004\\
03-04-06&2453828&1.264&17.0  $\pm$0.013&17.223$\pm$0.011&18.801$\pm$0.012&18.365$\pm$0.011\\
05-04-06&2453830&1.373&17.035$\pm$0.021&17.183$\pm$0.017&18.845$\pm$0.019&18.363$\pm$0.019\\
12-04-06&2453837&1.025&17.01 $\pm$0.013&17.174$\pm$0.011&18.782$\pm$0.013&18.348$\pm$0.012\\
13-04-06&2453838&1.166&17.0  $\pm$0.02 &17.189$\pm$0.017&18.783$\pm$0.019&18.315$\pm$0.018\\
15-04-06&2453840&0.884&16.985$\pm$0.006&17.213$\pm$0.005&18.77 $\pm$0.006&18.369$\pm$0.006\\
18-04-06&2453843&0.99 &16.986$\pm$0.005&17.205$\pm$0.004&18.771$\pm$0.005&18.355$\pm$0.005\\
25-04-06&2453850&0.918&16.987$\pm$0.004&17.197$\pm$0.004&18.77 $\pm$0.004&18.355$\pm$0.004\\
03-05-06&2453858&1.229&16.983$\pm$0.099&17.152$\pm$0.083&18.87 $\pm$0.094&18.386$\pm$0.092\\
10-05-06&2453865&0.914&16.976$\pm$0.007&17.204$\pm$0.006&18.761$\pm$0.006&18.351$\pm$0.006\\
13-05-06&2453868&0.942&16.976$\pm$0.006&17.191$\pm$0.005&18.755$\pm$0.006&18.349$\pm$0.006\\
17-05-06&2453872&1.093&16.965$\pm$0.007&17.205$\pm$0.006&18.732$\pm$0.007&18.344$\pm$0.007\\
20-05-06&2453875&1.438&17.068$\pm$0.056&17.093$\pm$0.047&18.734$\pm$0.053&18.394$\pm$0.052\\
23-05-06&2453878&1.428&16.956$\pm$0.025&17.228$\pm$0.02 &18.749$\pm$0.023&18.379$\pm$0.023\\
02-06-06&2453888&1.189&16.967$\pm$0.012&17.218$\pm$0.01 &18.748$\pm$0.011&18.34 $\pm$0.011\\
\hline
\end{longtable}

\end{document}